\documentclass[aps,floatfix,nofootinbib,preprint]{revtex4}
\usepackage{amsmath}
\usepackage{epsfig}
\usepackage{color}
\usepackage{endnotes}
\let\footnote=\endnote

\newcommand{\be}{\begin{equation}}
\newcommand{\ee}{\end{equation}}

\newcommand{\ber}{\begin{eqnarray}}
\newcommand{\eer}{\end{eqnarray}}

\newcommand{\bs}[1]{\ensuremath{\boldsymbol{#1}}}

\begin{document}

\title{Nonsymmetrized hyperspherical harmonics with realistic NN potentials}

\author{Sergio Deflorian$^{1,2}$,
  Nir Barnea$^{3}$,
  Winfried Leidemann$^{1,2}$, and
  Giuseppina Orlandini$^{1,2}$
  }

\affiliation{
  $^{1}$Dipartimento di Fisica, Universit\`a di Trento, I-38123 Trento, Italy \\
  $^{2}$Istituto Nazionale di Fisica Nucleare, Gruppo Collegato di Trento, I-38123 Trento, Italy \\
  $^{3}$Racah Institute of Physics, The Hebrew University, 91904, Jerusalem, Israel
}

\begin{abstract}
The Schr\"odinger equation is solved for an $A$-nucleon system using
an expansion of the wave function in nonsymmetrized hyperspherical harmonics. Our approach
is both an extension and a modification of the formalism developed
by Gattobigio {\it et al.}~\cite{GaK09,GaK11}. The extension consists in the inclusion of spin and isospin degrees
of freedom such that a calculation with more realistic NN potential models becomes possible, whereas
the modification allows a much simpler determination of the fermionic ground state.
The approach is applied to four- and six-body nuclei ($^4$He, $^6$Li) with various NN potential
models. It is shown that the results for ground-state energy and radius agree well with those
from the literature.
\end{abstract}

\bigskip


\maketitle

\section{Introduction} \label{sec:Intro}

Expansions of the wave function of an $A$-nucleon system on a specific basis set is a common tool
in few-nucleon ab initio calculations (for an overview see Ref.~\cite{LeO13}). One of the challenges
in such calculations is the implementation of the proper permutation symmetry, i.e.
of the antisymmetry for a fermion system like the atomic nucleus. A frequently used basis  
are the hyperspherical harmonics (HH). For a growing particle number an effective HH symmetrization 
method had been developed in Ref.~\cite{BaN98}. The method, however, requires considerable computational resources
with respect to memory and cpu time. Recently a different HH approach has been proposed in order
to construct wave functions with a proper permutation symmetry~\cite{GaK09,GaK11}. It consists in
the use of a nonsymmetrized HH (NSHH) basis thus avoiding the HH symmetrization procedure. 
Since the Hamiltonian commutes with the permutation operator,
all non degenerate eigenstates of the Hamiltonian have a well defined permutation symmetry.
The disadvantage is that additional effort has to be devoted to the determination of the respective symmetry.
On the other hand, in view of an application of the HH expansion method beyond $A=4$, the use of NSHH
might be advantageous.

In this work we extend the NSHH to consider spin and isospin degrees of freedom, 
which allows us to work with modern realistic NN potentials. In addition
we show how the antisymmetric ground state can be identified in a very simple and
effective way.
We apply our technique to calculate ground-state energies and radii
of $^4$He and $^6$Li with various NN potential models. In case of $^4$He we also use
the modern realistic AV18 potential~\cite{AV18}.

Our paper is organized as follows. The HH basis with inclusion of spin and isospin degrees of freedom is
discussed in Section~\ref{sec:HH}. The formalism of Ref.~\cite{GaK09,GaK11} for an efficient use of
a nonsymmetrized HH basis is laid out in the first part of Section~\ref{sec:NSHH}, whereas our modifications of 
the method are outlined in the second part of Section~\ref{sec:NSHH}. Our results for $^4$He and $^6$Li are discussed 
in Section~\ref{sec:Results}, which also includes a brief summary.

\section{The HH basis} \label{sec:HH}

To describe the HH basis, we first introduce the Jacobi coordinates. Here we use 
them in reversed order and for particles with equal mass,
\begin{equation}
\boldsymbol\eta_i=\sqrt{\frac{A-i}{A+1-i}}\left(\boldsymbol{r}_i-\frac{1}{A-i}\sum_{j=i+1}^A\boldsymbol{r}_j\right),
\ \ i=1,\ldots,N ,\,
\end{equation}
where $A$ is the number of particles, $N=A-1$, and $\boldsymbol{r}_i$ is the position vector 
for the $i$-th particle. To each Jacobi vector $\boldsymbol{\eta}_i$ 
an angular momentum operator $\bs{\hat{l}}_i$ is associated. For
the total orbital angular momentum operator up to the $n$-th coordinate ($n=2,\ldots,N$) one has 
\begin{equation}
\boldsymbol{\hat{L}}_n=\boldsymbol{\hat{L}}_{n-1}+\boldsymbol{\hat{l}}_n \,.
\end{equation} 
with $\boldsymbol{\hat{L}}_1 \equiv \boldsymbol{\hat{l}}_1$.
The hyperradial and the hyperangular 
variables $\rho_n$ and $\theta_n$ are defined as
\begin{equation}
\sin\theta_n=\frac{\eta_n}{\rho_n} \,, \ \ \rho_n=\sqrt{\sum_{i=1}^n\eta_i^2},\ \ n=2,\ldots,N \,.
\end{equation}
The $3N$ variables $\{\boldsymbol{\eta}_1,\ldots,\boldsymbol{\eta}_N\}$ are thus replaced by $\{\rho,\Omega_N\}$,
i.e. one hyperradial coordinate $\rho\equiv \rho_N$ and $3N-1$ hyperangular coordinates
$\Omega_N=\{\hat{\eta}_1,\ldots,\hat{\eta}_N,\theta_2,\ldots,\theta_N\}$.
The following relation,
\begin{equation}
\rho=\sqrt{\frac{1}{A}\sum_{j>i=1}^{A}\left(\boldsymbol{r}_j-\boldsymbol{r}_i\right)^2} \,,
\end{equation}
shows that the hyperradius is symmetric under permutation of particles.\\

Using the above set of coordinates, the Laplace operator for $n=$1,$\ldots,N$ can be rewritten as
\begin{eqnarray}
\Delta_n=\sum_{i=1}^n\nabla_{\eta_i}^2 & = & 
\frac{1}{\rho^{3n-1}}\frac{\partial}{\partial\rho}\rho^{3n-1}\frac{\partial}{\partial\rho}-
\frac{1}{\rho^2}\hat{K}_n^2 \nonumber \\
& = & \frac{\partial^2}{\partial\rho^2}+\frac{3n-1}{\rho}\frac{\partial}{\partial\rho}-
\frac{1}{\rho^2}\hat{K}_n^2 \,,
\end{eqnarray}
where $\hat{K}_n^2\left(\Omega_n\right)$ is a generalization of the usual angular momentum
operator $\hat{K}_1^2 \equiv \hat{l}_1^2 $ and is called the grand angular momentum operator. 
Its explicit expression for $n \ge 2$ is given by
\begin{eqnarray}
\hat{K}_n^2 & = & -\frac{\partial^2}{\partial\theta_n^2}+\frac{3n-6-(3n-2)\cos(2\theta_n)}
{\sin(2\theta_n)}\frac{\partial}{\partial\theta_n}\nonumber\\
& & +\frac{1}{\cos^2\theta_n}\hat{K}_{n-1}^2+\frac{1}{\sin^2\theta_n}\hat{l}_n^2 \,.
\end{eqnarray}

The hyperspherical harmonic functions $\mathcal{Y}_{\left[K\right]}\left(\Omega_N\right)$ 
are the eigenfunctions of the grand angular momentum operator:
\begin{equation}
\left[\hat{K}_N^2\left(\Omega_N\right)-K\left(K+3N-2\right)\right]\mathcal{Y}_{\left[K\right]}
\left(\Omega_N\right)=0 \,.
\end{equation}
The HH can be expressed in terms of the spherical harmonics $Y_{lm}(\hat{\eta})$ and of 
the Jacobi polynomials $P^{a,b}_\mu(z)$:
\begin{eqnarray}
\mathcal{Y}_{\left[K\right]}\left(\Omega_N\right) & = & \left[\sum_{m_1,\ldots,m_N}
\langle l_1m_1l_2m_2|L_2M_2\rangle \langle L_2M_2l_3m_3|L_3M_3 \rangle \ldots\right. \nonumber\\
& & \left.\times \langle L_{N-1}M_{N-1}l_Nm_N|L_NM_N \rangle \prod_{j=1}^NY_{l_jm_j}
\left(\hat{\eta}_j\right)\right] \nonumber\\
& & \times\left[\prod_{j=2}^N \mathcal{N}(K_j;l_jK_{j-1})(\sin\theta_j)^{l_j}(\cos\theta_j)
^{K_{j-1}}\right. \nonumber\\
& & \left.\times P_{\mu_j}^{[l_j+1/2],[K_{j-1}+(3j-5)/2]}\left(\cos\left(2\theta_j\right)\right)\right] \,.
\end{eqnarray}
The coefficients $\mathcal{N}_j(K_j;L_jK_{j-1})$ are normalization coefficients given by
\begin{equation}
\mathcal{N}_j(K_j;L_jK_{j-1})=\left[\frac{\mu_j!(2K_j+3j-2)\Gamma\left(\mu_j+K_{j-1}+l_j+
\frac{(3j-2)}{2}\right)}{\Gamma(\mu_j+l_j+\frac{3}{2})\Gamma\left(\mu_j+K_{j-1}+
\frac{(3j-2)}{2}\right)}\right]^{1/2},
\end{equation}
where the numbers $\mu_j$ are non-negative integers, and $K_j=K_{j-1}+2\mu_j+l_j$.

The basis functions consist not only of the hyperangular part, but also of 
the hyperradial part and of the spin and isospin part,
\begin{equation*}
|\Phi_i\rangle =|\mathcal{R}_{r_i}\mathcal{Y}_{[K]_i}\rangle\otimes
|\chi^{\rm spin}_{[S]_i}\chi^{\rm isospin}_{[T]_i}\rangle \,,
\end{equation*}
where $i$ enumerates the basis state $\Phi_i$, and $\mathcal{R}_{r}$ are the hyperradial 
functions numbered by the index $r$ (we use Laguerre polynomials). 
The hyperspherical harmonic functions $\mathcal{Y}_{[K]}$
are identified by the following set of quantum numbers 
\begin{equation}
[K]=\{K_N,K_{N-1},\ldots,K_{2},L_N,L_{N-1},\ldots,L_{2},l_N,l_{N-1},\ldots,l_{2},l_1\}
\end{equation}
and $\chi^{\rm spin}_{[S]}, \,\,\chi^{isospin}_{[T]}$ are the spin and isospin basis
states, respectively, defined by the sets of quantum numbers 
\begin{eqnarray}
[S] & = & \{S_A,S_{A-1},\ldots,S_2\}\\{}
[T] & = & \{T_A,T_{A-1},\ldots,T_2; T_{A,z}\}.
\end{eqnarray}
The spin quantum number $S_j$  
is the value of the coupled spin momenta of particles $1$ to $j$.
The isospin states are characterized in a similar way with the addition of
$T_{A,z}$, the third component of the total isospin $T_A$.

In order to consider potentials that depend on spin and isospin, one has to take 
as basis states the complete functions $|\Phi_i\rangle$, and so one has to 
consider the whole set of quantum numbers $\{[K],[S],[T]\}$ limited by a maximal
value $K_{\rm max}$ of $K_N$. Two 
possibilities arise: if the potential is central, then not only the total angular 
momentum $J$, the total isospin $T=T_A$ (isospin mixing neglected) and its projection $T_{A,z}$ 
are good quantum numbers for the eigenstates of $H$, 
but also the total orbital angular momentum $L=L_N$ and 
therefore the total spin $S=S_A$ (in fact $J$ becomes obsolete). 
In this case the total number of basis states 
is given by the number of radial functions times the number of hyperspherical harmonic
functions times the number of spin-isospin functions, subject to the condition that 
$S$, $L$, $T$ and $T_{A,z}$ have the desired values. On the other hand, if the potential is noncentral, 
like for example the realistic potential AV18, $L$ and $S$ are no longer good quantum 
numbers, leaving only $J$, $T$ and $T_{A,z}$ as constraints in the construction of the basis, 
and of course leading to a higher number of basis states. The size of the basis is the limiting 
factor in the applicability of the method. Examples for the 
number of basis states are presented in Section~\ref{sec:Results}.

\section{Expansions with nonsymmetrized HH functions} \label{sec:NSHH}

In order to describe the nonsymmetrized HH (NSHH) method
introduced by Gattobigio {\it et al.}~\cite{GaK09,GaK11},
we first discuss a few aspects concerning permutation symmetry.
The HH functions defined above do not have well-defined permutational symmetries. 
A particle permutation  changes the definition of the Jacobi coordinates 
as well as the couplings between the different angular momenta. Consequently,
the effect of particle permutation on the HH functions is rather complicated, and the matrices 
representing the permutation operators $\hat{P}_{ij}$ on the HH basis are arbitrary 
matrices (not diagonal or block-diagonal matrices), and have to be calculated numerically.
In order to reduce the evaluation of the matrix elements of the
  potential acting between particles $i$ and $j$ into a one dimensional
  integral one has to use Jacobi vectors 
  such that $\bs{\eta}_{A-1}=\sqrt{\frac{1}{2}}(\bs{r}_{i}-\bs{r}_j)$. This
  transformation can be realized through the permutations $\hat P_{i,A-1}$ and
  $\hat P_{j,A}$ acting on the HH basis states.
The matrix representing the permutation operator $\hat P_{ij}$ can be written as
\begin{equation}
\mathcal{B}^{ij}_{[K,S,T][K^\prime,S^\prime,T^\prime]}=\langle \Phi_{[K,S,T]}
(\Omega^{ij}_N)|\Phi_{[K^\prime,S^\prime,T^\prime]}(\Omega_N)\rangle \,,
\end{equation}
where the set of quantum numbers $[K,S,T]\equiv\{[K],[S],[T]\}$ identify 
the orbital, spin and isospin quantum numbers of the basis functions, 
$\Omega_N$ is the set of all hyperangular variables defined above and 
$\Omega^{ij}_N$ indicates the set of hyperangular variables with the particles 
$i$ and $j$ interchanged. All permutations can be expressed as a product of 
a certain number $\nu_{ij}$ of transpositions (depending on the particles indexes 
$i$ and $j$) in which only adjacent particles are interchanged,
\begin{equation}
\hat{P}_{ij}=\prod_{l=1}^{l=\nu_{ij}}\hat{P}_{k_l}=\hat{P}_{k_1}\ldots\hat{P}_{k_{\nu_{ij}}} \,,
\end{equation}
where the operator $\hat{P}_k$ exchanges particles $k$ and $k+1$. This can be 
interpreted as follows: each particle carries a number and occupies originally a
a box with the same number. To bring particle number $i$ in the box $j$ and 
vice versa, one can act with a certain number $\nu_{ij}$ of permutations 
that exchange only particles in adjacent boxes. Thanks to the properties of
the HHs, the matrices 
\begin{equation}
\mathcal{B}^{k,k+1}_{[K,S,T][K^\prime,S^\prime,T^\prime]}=\langle\Phi_{[K,S,T]}(\Omega^{k,k+1})
|\Phi_{[K^\prime,S^\prime,T^\prime]}(\Omega_N)\rangle
\end{equation}
representing the operators $\hat{P}_k$ are block diagonal and can be easily calculated. 
This provides an easy and fast way to calculate the product of the $\mathcal{B}$ matrices 
 on state vectors.

We remark that the use of the NSHH has been developed in order to avoid the need for 
the symmetrization of the basis functions, which, as already pointed out in the introduction, 
requires non negligible computational resources. 
Though the number of basis functions for equal values of $K_{\rm max}$ is in general considerably 
larger in the NSHH method than with symmetrized functions, 
it might be still advantageous to use nonsymmetrized functions.

In Ref.~\cite{GaK11} bound states have been calculated 
using the Volkov potential~\cite{volkov1965}, which is central and independent on spin and isospin. 
In this case the spatial wave function can be 
treated separately  from the spin-isospin part. The Hamiltonian is represented 
on the HH and  diagonalized. Then the symmetry of the eigenstates is analyzed 
and the spin-isospin part with the correct symmetry is multiplied in order to obtain 
an antisymmetric wave function. It is clear that following this procedure the number 
of HH basis states, 
and therefore the size of the Hamiltonian matrix to be diagonalized,
is relatively low. 
However, this procedure only applies to  potentials that do not depend 
on spin and isospin. 

In the following we describe how to modify the method outlined above 
in order to be able to treat also realistic potentials.
In this case we need to deal with basis functions in the J-coupling
as described in Section~\ref{sec:HH}.  This leads to an increase 
in the size of the Hamiltonian matrix and therefore in the number of eigenfunctions 
that need to be analyzed in terms of symmetry.
In order to speed up the search for the antisymmetric ground state  
we have implemented a method 
based on the use of the transposition class sum operator, the  Casimir operator, of the permutation group.
This method is  analogous to the Lawson method~\cite{GoL74} for the removal 
of the spurious center of mass motion in Shell Model calculations.

The Casimir operator of the permutation group $\hat{C}(A)=\sum_{j>i=1}^A\hat{P}_{ij}$ 
and the Hamiltonian $H$ commute; hence they can be diagonalized
  simultaneously.
In general, the eigenvalues of the operator $\hat C(A)$ are not sufficient
  to identify the irreducible representations of the permutation
  group. However, the completely symmetric and antisymmetric representations
  correspond to the extreme eigenvalues of $\hat C(A)$ and are well separated 
  from the rest of the spectrum. Antisymmetric states correspond to the lowest
  eigenvalue of $\hat{C}(A)$,  
  while symmetric states correspond to the highest one, in detail one has
\begin{flalign}
\hat{C}(A) \Psi_S & = \frac{A(A-1)}{2} \Psi_S =\lambda_S \Psi_S \,,\nonumber\\
\hat{C}(A) \Psi_M & = \lambda_M \Psi_M \,, \nonumber\\
\hat{C}(A) \Psi_A & = -\frac{A(A-1)}{2} \Psi_A =\lambda_A \Psi_A \,,
\end{flalign}
where $\Psi_S$, $\Psi_M$, $\Psi_A$, correspond to symmetric, 
mixed-symmetry, and antisymmetric states, respectively, and $\lambda_A < \lambda_M <\lambda_S$.
We diagonalize the matrix
\begin{equation}
H^\prime=H+\gamma \hat{C}(A) \,,
\end{equation}
where $\gamma$ is a real parameter. The eigenvalues of $H^\prime$ are given by
\begin{equation}
E^\prime_{n,\Gamma}=E_{n,\Gamma}+\gamma\lambda_\Gamma \,, 
\end{equation}
where $E_{n,\Gamma}$ ($n=0,1,2,\ldots, N_{\rm max}(\Gamma)$) are the eigenvalues of $H$ 
for the symmetry $\Gamma$ ($\Gamma=S,M,A$). 
We denote by $\cal{E}$ the lowest eigenvalue of the Hamiltonian $H$ (${\cal E} = {\rm min} \{E_{0,S},E_{0,M},E_{0,A}\}$),
which can be found by performing the calculation with $\gamma=0$.

To calculate the lowest antisymmetric eigenvalue of $H$, i.e. $E_{0,A}$, 
we choose $\gamma>0$ large enough so that  $E^\prime_{0,A}$ is by far 
the lowest eigenvalue of $H^\prime$. Thus one imposes the relations
\begin{eqnarray}
E^\prime_{0,A}=E_{0,A}+\gamma\lambda_A<E_{n,\Gamma}+\gamma\lambda_\Gamma\ \ \ \forall n,\ \ \forall \Gamma=S,M \,,
\end{eqnarray}
which corresponds to
\begin{equation}
\gamma>\frac{E_{0,A}-E_{n,\Gamma}}{\lambda_\Gamma-\lambda_A} \,.
\end{equation}
Assuming that $E_{0,A}<0$, and using the fact that $\lambda_\Gamma-\lambda_A \ge A$,
the following condition is sufficient
\begin{equation} \label{condition}
  \gamma > {\frac{|{\cal E}|}{A}} \,.
\end{equation}
Thus, with a proper value of $\gamma$ the lowest eigenstate of the
  Hamiltonian matrix $H'$ is the physical antisymmetric wave-function, 
and the correct value of the ground-state energy $E_{0,A}$ is obtained by subtracting 
  $\gamma\lambda_A$ from $E^\prime_{0,A}$. Of course with a proper choice of
  $\gamma$ one can also calculate excited states.
To this end, there is no need to calculate more than a few eigenstates 
of the Hamiltonian matrix, and one can use the Lanczos algorithm.

\section{Discussion of results} \label{sec:Results}
Within the present formalism we have calculated ground-state energies and radii of $^4$He and $^6$Li with
various NN potentials using the following models: Volkov~\cite{volkov1965}
(central), MTI/III~\cite{MaT69} and Minnesota (MN)~\cite{MN} (central spin-isospin dependent);  in addition
for $^4$He we have used AV4$^\prime$~\cite{WiP02} (central spin-isospin dependent) and the realistic AV18~\cite{AV18}.
In all our calculations the Coulomb force is included and the isospin mixing is neglected.
Since for Volkov, MN, and MTI/III interactions there is no unique parameter setting we define these potential
models in Table~\ref{tab:pots}.
\begin{table}
\caption{Definition and parameter setting for Volkov, MN, and MTI/III  NN potentials.
Volkov: $V(r)=V_1\,$exp$[-(r/\mu_1)^2]+V_2\,$exp$[-(r/\mu_2)^2]$;
MN: $V(r)=V_1\,$exp$(-\mu_1r^2)+V_2\,$exp$(-\mu_2r^2)$ and $V(r)=V_1\,$exp$(-\mu_1r^2)+V_3\,$exp$(-\mu_3r^2)$ 
for NN channels with spin $S=0$, isospin $T=1$ and $S=1$, $T=0$, respectively;
MTI/III: $V(r)=V_1\,$exp$(-\mu_1r)/r+V_2\,$exp$(-\mu_2r)/r$ ($S=0,$ $T=1$),
$V(r)=V_1\,$exp$(-\mu_1r)/r+V_3\,$exp$(-\mu_3r)/r$ ($S=1,$ $T=0$). The parameters $V_i$ are in units of MeV (Volkov, MN) and of MeV$\cdot$fm (MTI/III); the parameters $\mu_i$ are in units of fm, fm$^{-2}$, and fm$^{-1}$ for Volkov, MN, and
MTI/III potential, respectively.}\label{tab:pots}
\begin{center}
\begin{tabular}{c|cc|cc|cc} \hline\hline
 & \multicolumn{2}{c}{Volkov}\vline& \multicolumn{2}{c}{MN} \vline & \multicolumn{2}{c}{MTI/III}\\
\hline
    i  &    $V_i$  &     $\mu_i$  &    $V_i$  &     $\mu_i$  &    $V_i$  &     $\mu_i$ \\
\hline
$\;\;\;$1$\;\;\;$  & $\;\;$144.86$\;\;$ & $\;\;$0.82$\;\;$ & $\;\;\;$200.0$\;\;\;$ & $\;\;\;$1.487$\;\;\;$ & $\;\;\;$1458.047$\;\;\;$ & $\;\;\;$3.11$\;\;\;$\\
$\;\;\;$2$\;\;\;$  & $\;\;$--83.34$\;\;$ & $\;\;$1.60$\;\;$ & $\;\;\;$--91.85$\;\;\;$ & $\;\;\;$0.465$\;\;\;$ & $\;\;\;$--520.872$\;\;\;$ & $\;\;\;$1.555$\;\;\;$\\
$\;\;\;$3$\;\;\;$  & $\;\;\;$--$\;\;\;$ & $\;\;\;$--$\;\;\;$ & $\;\;\;$--178.0$\;\;\;$ & $\;\;\;$0.639$\;\;\;$ & $\;\;\;$--635.306$\;\;\;$ & $\;\;\;$1.555$\;\;\;$\\
\hline
\hline
\end{tabular}
\end{center}
\end{table}

To illustrate the typical size of the HH basis, in Table~\ref{tab:basis} we list its dimension
for increasing $K_{\rm max}$. It is evident that the number of basis functions grows
very rapidly.
Comparing for example the dimensions with $K_{\rm max}=10$, one sees that for $^4$He the number of HH states grows from
about 1000 for a central interaction to about 5000 for a noncentral potential. For $^6$Li one already starts with about
500000 states with a central force which is then increased to about 7 million states with a noncentral potential.
Including the hyperradial part the size of the Hamiltonian matrix increases further. In fact
the numbers in the table have to be multiplied by a factor 20 which is the number of Laguerre polynomials
that we consider for each $K$.
\begin{table}
\caption{Number of NSHH basis functions (including spin-isospin functions, but without hyperradial part) for increasing 
values of $K_{\rm max}$, for $^4$He and $^6$Li with central and noncentral potentials.}\label{tab:basis}
\begin{center}
\begin{tabular}{c|cc|cc} \hline\hline
 & \multicolumn{2}{c}{$^4$He} \vline & \multicolumn{2}{c}{$^6$Li} \\
\hline
$K_{\rm max}$  & central  & noncentral  & central  & noncentral   \\
\hline
    2  &    24  &     54  &     675  &  2750\\
    4  &    84  &    264  &    5400  &  40025\\
    6  &   224  &    852  &   30600  &  315675\\
    8  &   504  &   2172  &  137025  &  1728950\\
   10  &  1008  &   4746  &  514215  &  7392960\\
   12    &  1848  &   9269  & 1678950  & 26377350 \\
   14  &  3168  &  16776 & -- & -- \\
   16  &  5148  &  28404 & -- & -- \\
   18  &  8008  &  45694 & -- & -- \\
   20  & 12012  &  79488 & -- & -- \\
   22  & 17472  & 104988 & -- & -- \\
   24  & 24752  & 151788 & -- & -- \\
\hline\hline
\end{tabular}
\end{center}
\end{table}

To improve the convergence we have used the EIHH (Effective Interaction Hyperspherical Harmonics) formalism \cite{BaL00}.
To distinguish between the calculations without and with effective interaction we use the acronyms NSHH and EI-NSHH,
respectively.
The use of effective interaction accelerates the convergence considerably, as can be seen in Table~\ref{tab:He4_volkov_mt}, 
where we list $^4$He ground-state energies and root mean square (RMS) radii obtained  with the Volkov and the MTI/III potentials. One notes  that the use of an effective interaction is not very important for the Volkov potential, whereas
it has a great effect in case of the MTI/III interaction. Considering for example the low value of $K_{\rm max}=4$ one 
obtains already results close to the converged ones for all cases, except for the MTI/III without effective interaction. 
The difference is explained by the fact that the Volkov force has a very soft core, whereas the MTI/III potential has a rather
strong short-range repulsion, which requires HH functions with a rather large $K_N$. Thus, in order to accelerate the 
convergence, we use the EIHH method for all the results discussed further below. In the table we list in addition a 
selection of results from the literature (note that for the MTI/III potential parameter settings different from ours are
also in use). Comparing with our NSHH results one observes a very good agreement.
\begin{table}
\caption{$^4$He ground-state energy $E_0$ and radius $r_{\rm RMS}$ 
in units of MeV and fm, respectively, with Volkov and MTI/III potentials with (EI-NSHH) and without (NSHH) effective 
interaction.}\label{tab:He4_volkov_mt}
\begin{center}
\begin{tabular}{ccccc|cccc} \hline\hline
 & \multicolumn{4}{c} {Volkov} \vline & \multicolumn{4}{c}{MTI/III}\\
\hline
 & \multicolumn{2}{c}{NSHH} & \multicolumn{2}{c}{EI-NSHH} \vline& \multicolumn{2}{c}{NSHH} & \multicolumn{2}{c}{EI-NSHH}\\
\hline
$K_{\rm max}$ \vline & $\;\;\;E_0\;\;\;$  & $\;\;\;r_{\rm RMS}\;\;\;\;\;$  & $\;\;\;E_0\;\;\;$  & $\;\;\;r_{\rm RMS}\;\;\;$  & $\;\;\;E_0\;\;\;$  & $\;\;\;\;\;r_{\rm RMS}\;\;\;$  & $\;\;\;E_0\;\;\;$  & $\;\;\;r_{\rm RMS}\;\;\;$ \\
\hline
 $\;\;\;\,$2$\;\;\,\,$ \vline &  --28.579  &   1.4877 &      --    & --         & $\;\;$--        & --         & --         &  -- \\
 $\;\;\;\,$4$\;\;\,\,$ \vline &  --29.281  &   1.4920 &  --30.414  &   1.4907   & $\;\;$--9.679   &   1.8667   & --30.782   &  1.4236\\
 $\;\;\;\,$6$\;\;\,\,$ \vline &  --29.811  &   1.4867 &  --30.417  &   1.4905   & $\;\;$--13.744  &   1.6401   & --30.599   &  1.4263\\
 $\;\;\;\,$8$\;\;\,\,$ \vline &  --30.160  &   1.4847 &  --30.449  &   1.4901   & $\;\;$--19.923  &   1.4612   & --30.922   &  1.4193\\
    $\;\;$10$\;\;\;$ \vline &   --30.276  &   1.4854 &   --30.407  &   1.4899   & $\;\;$--22.807  &   1.4196   & --30.584   &  1.4235\\
    $\;\;$12$\;\;\;$ \vline  &  --30.363  &   1.4871 &   --30.422  &   1.4902   & $\;\;$--25.808  &   1.3984   & --30.763   &  1.4221\\
    $\;\;$14$\;\;\;$ \vline  &  --30.390  &   1.4881 &   --30.416  &   1.4900   & $\;\;$--27.253  &   1.3938   & --30.679   &  1.4223\\
    $\;\;$16$\;\;\;$ \vline  &  --30.405  &   1.4889 &   --30.417  &   1.4900   & $\;\;$--28.385  &   1.3946   & --30.700   &  1.4221\\
    $\;\;$18$\;\;\;$ \vline  & -- & -- & -- & -- & $\;\;$--29.061  &   1.3978   &  --30.687   &  1.4221\\
    $\;\;$20$\;\;\;$ \vline  & -- & -- & -- & -- & $\;\;$--29.558  &   1.4015   &  --30.696   &  1.4220\\
    $\;\;$22$\;\;\;$ \vline  & -- & -- & -- & -- & $\;\;$ --29.856  &   1.4048   &  --30.687   &  1.4219\\
    $\;\;$24$\;\;\;$ \vline  & -- & -- & -- & -- & $\;\;$--30.092  &   1.4081   &  --30.693   &  1.4220\\
\hline
NSHH \cite{GaK11} & & & --30.418 & \\
HH \cite{viviani2005,kievsky2008} & & & --30.420 & 1.490\\
SVM \cite{SVM} & & & --30.42 & & & & & \\
EIHH \cite{BaL00} & & & & & & & --30.71 & 1.4222\\
CHH \cite{BaE01} & & & & & & & --30.69 & 1.421\\


\hline\hline
\end{tabular}
\end{center}
\end{table}

In Table~\ref{tab:He4_minnesota_AV4p} we list results for  $^4$He obtained with MN and AV4$^\prime$ potentials. Also in this case
we find a very good convergence and a good agreement with results from other authors.
As already shown in Table~\ref{tab:basis} the number of HH states increases significantly using a realistic NN force instead of
a central potential model. Nonetheless, as Table~\ref{tab:He4_AV18} illustrates, we are also
able to reach convergent results for $^4$He with the modern realistic AV18 potential. In the table we also make a detailed 
comparison with an EIHH calculation with symmetrized HH states. It is evident that both calculations lead essentially to the
same results. Further results from other groups displayed in Table~\ref{tab:He4_AV18} also confirm the high precision 
of present-day few-nucleon calculations. 
\begin{table}
\caption{EI-NSHH results for $E_0$ and $r_{\rm RMS}$ (units as in Table~\ref{tab:He4_volkov_mt}) of $^4$He with MN and AV4$^\prime$ potentials.}\label{tab:He4_minnesota_AV4p}
\begin{center}
\begin{tabular}{c|cc|cc} \hline\hline
 & \multicolumn{2}{c}{MN} \vline & \multicolumn{2}{c}{AV4$^\prime$}\\
\hline
$K_{\rm max}$ & $\;\;\;E_0\;\;\;$  & $\;\;\;\;\;r_{\rm RMS}\;\;\;\;\;$  & $\;\;\;\;\;\;E_0\;\;\;\;\;\;$  & $\;\;\;r_{\rm RMS}\;\;\;$ \\
\hline
   2  &\; --29.723 & 1.4049  &  --32.258  &   1.3698\\
   4  &\; --30.065 & 1.4139  &  --32.227  &   1.3840\\
   6  &\; --29.950 & 1.4112  &  --31.781  &   1.3957\\
   8  &\; --29.981 & 1.4108  &  --32.579  &   1.3798\\
  10  &\; --29.937 & 1.4104  &  --31.858  &   1.3883\\
  12  &\; --29.951 & 1.4107  &  --32.201  &   1.3866\\
  14  &\; --29.945 & 1.4104  &  --32.047  &   1.3865\\
  16  &\; --29.946 & 1.4104  &  --32.068  &   1.3865\\
  18  &\; --29.945 & 1.4104  &  --32.051  &   1.3865\\
  20  &\; --29.945 & 1.4104  &  --32.060  &   1.3865\\
  22  &\; --29.945 & 1.4104  &  --32.049  &   1.3862\\
  24  &\; --29.945 & 1.4103  &  --32.054  &   1.3863\\
\hline
EIHH \cite{BaL00} & --29.96 & 1.4106 & &\\
HH \cite{viviani2005,kievsky2008} &\; --29.947 & 1.4105 & &\\
SVM \cite{SVM} &\; --29.937 & & &\\

GFMC \cite{WiP02} & & &\; --32.11(2) &\\
\hline\hline
\end{tabular}
\end{center}
\end{table}
\begin{table}
\caption{EI-NSHH results for $E_0$ and $r_{\rm RMS}$ (units as in Table~\ref{tab:He4_volkov_mt}) of $^4$He with AV18 potential.}\label{tab:He4_AV18}
\begin{center}
\begin{tabular}{c|cc|cc} \hline\hline
 & \multicolumn{2}{c}{present work} \vline & \multicolumn{2}{c}{$\;\;$Reference \cite{gazit2006}}\\
\hline
$K_{\rm max}$ & $\;\;\;\;\;\;E_0\;\;$ & $\;\;r_{\rm RMS}\;\;$ & $\;\;E_0\;\;$ & $\;\;r_{\rm RMS}\;\;$\\
\hline
   2  &   $\;\;$--24.640  &   1.5063 & -- & -- \\
   4  &   $\;\;$--26.124  &   1.5111 & -- & -- \\
   6  &   $\;\;$--25.311  &   1.5061 & $\;\;$--25.312 & 1.506\\
   8  &   $\;\;$--24.999  &   1.5089 & $\;\;$--25.000 & 1.509\\
  10  &   $\;\;$--24.442  &   1.5197 & $\;\;$--24.443 & 1.520\\
  12  &   $\;\;$--24.491  &   1.5176 & $\;\;$--24.492 & 1.518\\
  14  &   $\;\;$--24.348  &   1.5184 & $\;\;$--24.350 & 1.518\\
  16  &   $\;\;$--24.313  &   1.5181 & $\;\;$--24.315 & 1.518\\
  18  &   $\;\;$--24.271  &   1.5177 & $\;\;$--24.273 & 1.518\\
  20  &   $\;\;$--24.266  &   1.5176 & $\;\;$--24.268 & 1.518\\
  22  &   $\;\;$--24.246  &   1.5170 & -- & -- \\
\hline
HH \cite{viviani2005,kievsky2008} & --24.22 & 1.512 & &\\
FY \cite{NoK02} & --24.23 & & &\\
FY \cite{LaC04} & --24.22 & 1.516 & &\\
AGS \cite{DeF07} & --24.24 & & & \\
\hline\hline
\end{tabular}
\end{center}
\end{table}

Now we turn to the $^6$Li nucleus. In Table~\ref{tab:Li6_various} we present results for central potential models only.
The reason is the non-parallel character of our present code, which prevents us from using a very large basis
(in future we plan to work with a parallel code which should allow to obtain converged results for $A=6$-8 with more 
realistic nuclear force models).
The Table shows that, similarly as for $^4$He case, the results for the Volkov potential converge already for rather low 
$K_{\rm max}$ values. On the contrary for the MN interaction
and even more for the MTI/III potential it would be desirable to make calculations for even higher $K_{\rm max}$ values
in order to obtain a more converged result for the radius. However, our $^6$Li results are of the same
level of precision as those from the literature and the agreement is satisfying.
\begin{table}
\caption{EI-NSHH results for $E_0$ and $r_{\rm RMS}$ (units as in Table~\ref{tab:He4_volkov_mt}) of $^6$Li with various potentials.}\label{tab:Li6_various}
\begin{center}
\begin{tabular}{c|cc|cc|cc} \hline\hline
 & \multicolumn{2}{c}{Volkov} \vline& \multicolumn{2}{c}{$\;\;\;\;\;\;\;\;$MTI/III$\;\;\;\;\;\;\;\;$} \vline & \multicolumn{2}{c}{MN}\\
\hline
$K_{\rm max}$ & $E_0$  & $\;\;\;\;r_{\rm RMS}\;\;\;\;$  & $E_0$ & $r_{\rm RMS}$ & $E_0$  & $\;\;\;\;r_{\rm RMS}\;\;\;\;$ \\
\hline
    2   &  $\;\;$--75.468  &   1.6823   &  $\;\;$--46.991  &   2.1609   &  $\;\;$--47.402  &   2.1297 \\
    4   &  $\;\;$--66.525  &   1.5922   &  $\;\;$--36.689  &   2.0793   &  $\;\;$--35.712  &   2.0813  \\
    6   &  $\;\;$--66.725  &   1.6005   &  $\;\;$--36.450  &   2.1202  &  $\;\;$--35.277  &   2.1298  \\
    8   &  $\;\;$--66.577  &   1.5993   &  $\;\;$--36.380  &   2.1380  & $\;\;$--35.067 & 2.1504\\
   10   &     --      &   --         &  $\;\;$--36.324  &   2.1562 &  -- & -- \\
   12   &     --      &   --         &  $\;\;$--36.269  &   2.1796 & -- & -- \\
\hline
NSHH \cite{GaK11} & --66.491 & & & \\
EIHH \cite{BaL00} & & & --36.6 & 2.15 & --35.2 & 2.1\\
HH \cite{barnea1999} & --66.57 & & --35.91 & \\
\hline\hline
\end{tabular}
\end{center}
\end{table}

We summarize our work as follows. The Schr\"odinger equation is solved for an $A$-body nucleus using
an expansion of the wave function in nonsymmetrized hyperspherical harmonics. Our approach
is both an extension and a modification of the formalism developed
by Gattobigio {\it et al.}~\cite{GaK09,GaK11}. The extension consists in the inclusion of spin and isospin degrees
of freedom such that a calculation with more realistic NN potential models becomes possible.
Different from Refs.~\cite{GaK09,GaK11} the cumbersome application of projection operators in order to determine
the permutation symmetry of the various eigenfunctions of the Hamiltonian $H$ is avoided. Instead,  a potential term
proportional to the Casimir operator ${\hat C}(A)$ is added, such that, after diagonalizing the new Hamiltonian,
the state with the lowest energy is manifestly antisymmetric.
Since one knows the eigenvalue of the additional potential term, and since $H$ and ${\hat C}$ commute
the ground-state energy and wave function of the original $A$-body problem can be obtained in a straightforward manner.
In this work the approach has been used to calculate binding energy and radii of four- and six-body nuclei ($^4$He, $^6$Li) 
with various NN potential models and it has been shown that the results agree with those from the literature.

\section*{Acknowledgements}
The work of N.B. was supported by the Israel Science Foundation (Grant number 954/09). 
W.L. and G.O. acknowledge support from the MIUR grant PRIN-2009TWL3MX.

\end{document}